# A linear approximation algorithm for the BPP with the best possible absolute approximation ratio

Abdolahad Noori Zehmakan*, ETH University, abdolahad.noori@inf.ethz.ch

Mojtaba Eslahi, Allameh Tabatabai University, eslahikelorazi921@atu.ac.ir

Abstract:

*The Bin Packing Problem is one of the most important Combinatorial Optimization problems in optimization and has a lot of real-world applications. Many approximation algorithms have been presented for this problem because of its NP-hard nature. In this article also a new creative approximation algorithm is presented for this important problem.*

*It has been proven that the best approximation ratio and the best time order for the Bin Packing Problem are $\frac{3}{2}$ and $O(n)$, respectively unless $P = NP$. The presented algorithm in this article has the best possible factors, $O(n)$ and $\frac{3}{2}$.*

## 1. Introduction

The Bin Packing Problem (BPP) is an optimization problem which like lots of interesting optimization problems is NP-hard. In recent years, researchers have made several attempts to solve these problems with approximation approaches [1, 2, 3]. In the approximation algorithms two parameters are important, time order and approximation ratio. An $\alpha$-approximation algorithm for an optimization problem is a polynomial time algorithm that for all instances of the problem produces a solution which its value is within a factor of $\alpha$ of the value of an optimal solution. For an $\alpha$-approximation algorithm, $\alpha$ will be called the *approximation ratio* of the algorithm. In the literature, it is also called the performance guarantee or approximation factor of the algorithm. Generally, for values of $\alpha > 1$ minimization problems, and for values of $\alpha < 1$ maximization problems are followed. (When $\alpha$ is bigger than 1 it means it is related to a minimization problem, otherwise it is relevant to a maximization problem) Thus a ½-approximation algorithm for a maximization problem is a polynomial-time algorithm that always returns a solution that its value is at least half the optimal value [4]. The size of the computed solution and the size of the optimal solution are shown by $P^*$ and P respectively. Based on the definition it is concluded that $|P^*| \leq \alpha |P|$, however for the maximum optimization problems $|P^*| \geq \alpha |P|$ applies.

The problem has many real-world applications like in stock-cutting, loading trucks, railway carriages, and others, but it is NP-hard; therefore in the last decades a lot of approximation algorithms have been developed to compute near-optimal solutions since in practice near-optimally is considered to be often acceptable enough. Almost all the suggested approximation algorithms for this problem use classical methods like Greedy, Dynamic Programming [6] and Rounding, Deterministic Rounding, Random

---

*Corresponding Author.

Email address: abdolahad.noori@inf.ethz.ch and ahad_ie2000@yahoo.com , Tel: +41446324329.

Sampling & Randomized Rounding, Primal Dual [9], but in this article a creative approximation algorithm for solving the Bin Packing Problem will be used.

In the Bin Packing Problem, n items were given with specified weights $a_1, a_2, \ldots, a_n$ such that:

$$0 < a_1 \leq a_2 \leq \cdots \leq a_n < 1$$

With the aim of Packing the objects in the minimum number of bins, and ensuring that the sum of the objects' weights does not exceed the bin capacity in any bin.

The Bin Packing Problem is related to a decision problem called the Partition Problem. In the Partition Problem, we are given n positive integers $b_1, b_2, \ldots, b_n$ whose sum B=$\sum_{i=1}^{n} b_i$ is even, and the purpose is to know if the partition of the set of indices {1, 2, …, n} into sets and T that $\sum_{i \in S} b_i = \sum_{i \in T} b_i$, is possible. The partition problem is well-known to be NP-Complete [4].

One of the most famous approximation algorithms for the BPP is the First-Fit-Decreasing algorithm (FFD), where the pieces are packed in order of non-decreasing size, the next piece is always packed into the first bin in which it fits; that is when bin 1 is first opened, and starting on bin k+1 is only started when the current piece does not fit into any of the bins 1, …, k. this method is one of the most well-known methods for Bin Packing Problem and has many different versions [10, 11, 7].

Simchi-Levi [8] proved that FFD and BFD algorithms have an absolute worst-case ratio of 3/2. These algorithms run in time $O(n \log n)$. Zhang and Xiaoqiang [2] provided a linear time constant space off-line approximation algorithm with absolute worst-case ratio 3/2 and they also presented a linear time constant space on-line algorithm and prove that the absolute worst-case ratio is 3/2. In 2003, Rudolf and Florian [5] presented an approximation algorithm for the bin packing problem which has a linear running time and absolute approximation factor of 3/2. Moreover, Noori Zehmakan [1] also presents two heuristic approximation algorithms. The first one is a constant-space $\frac{3}{2}$-approximation algorithm, and the second one is a modified linear version of FFD. He also shows that these two algorithms not only enjoy best possible theoretical criteria, but also perform much better than some other popular and efficient algorithms like FFD, Guochuan [2], and Rudolf [5] by using experimental results on standard database of OR-LIBRARY.

Many different versions of the Bin Packing Problem have been presented. For instance, in [12] a 13/12 approximation algorithm for bin packing with extendable bins has been presented. The bin packing problem with item fragmentation also has been discussed [13] such that the items may be fragmented at a price. In addition, asymptotic fully polynomial approximation schemes for variants of open –end bin packing problem have been presented [14].

As mentioned before, it has been proven that the best algorithm for Bin Packing Problem has the approximation ratio of 3/2 and the time order of $O(n)$ unless $P = NP$ [8].The presented algorithm in this article does not follow any of the mentioned approaches, and it is a heuristic algorithm. This inventive approach and the new way of solving BPP have even been expanded into the step of proof. The method of proof can be a strong pattern for proving and achieving the approximation ratio of the approximation algorithms.

The reminder of the paper is organized as follows. In Section 2, the suggested algorithm based on the ranging is present and after that in Section 3, a heuristic and long proof is discussed for the approximation ratio of the algorithm. In Sections 4 and 5, the idea of creating more ranges and the scaling method for the improving the suggested algorithm are explained.

## 2. The Suggested Algorithm

The inputs are separated into 2 different groups, *Small Items*(S) and *Large Items* (L). The main idea of the algorithm is to put the inputs into 10 equal ranges and match the complementary ranges. The algorithm attempts to match the S and the L in an intelligent manner. This procedure is then continued until the L items are finished. After that, if there is any remaining S item, they will be matched with each other. This procedure is repeated until all S item are processed.

The suggested algorithm:

1- Read $n$ inputs $a_1, a_2, \ldots, an$.

2- Classify the inputs under 10 equal size ranges $(0, 0.1), (0.1, 0.2) \ldots (0.9, 1)$.

3- If there are not any items in $(0.5, 0.6)$ go to line 11.

4- Initialize the pointer $a$ with one of the items from $(0.5, 0.6)$ randomly.

5- If there is at least one item in $(0.4, 0.5)$, initialize pointer $b$ with an item from it randomly. Set $c = a + b$. If $c \leq 1$, put it in the appropriate range as a new item and remove $a$ and $b$ from their ranges and Go to line 3.

6- If there is at least one item in $(0.3, 0.4)$, initialize pointer $b$ with an item from it randomly. Set $c = a + b$. If $c \leq 1$, put it in the appropriate range as a new item and remove $a$ and $b$ from their ranges. Go to line 3.

7- If there is at least one item in $(0.2, 0.3)$, initialize pointer $b$ with an item from it randomly. Set $c = a + b$. If $c \leq 1$, put it in the appropriate range as a new item and remove $a$ and $b$ from their ranges. Go to line 3.

8- If there is at least one item in $(0.1, 0.2)$, initialize pointer $b$ with an item from it randomly. Set $c = a + b$. If $c \leq 1$, put it in the appropriate range as a new item and remove $a$ and $b$ from their ranges. Go to line 3.

9- If there is at least one item in $(0, 0.1)$, initialize pointer $b$ with an item from it randomly. Set $c = a + b$. If $c \leq 1$, put it in the appropriate range as a new item and remove $a$ and $b$ from their ranges. Go to line 3.

10- Put $a$ in a new bin, remove it from its range and go to line 3.

11- If there are not any items in $(0.6, 0.7)$ go to line 18.

12- Initialize pointer $a$ with one of the items from $(0.6, 0.7)$ randomly.

13- If there is at least one item in $(0.3, 0.4)$, initialize pointer $b$ with an item from it randomly. Set $c = a + b$. If $c \leq 1$, put it in the appropriate range as a new item and remove $a$ and $b$ from their ranges. Go to line 11.

14- If there is at least one item in $(0.2, 0.3)$, initialize pointer $b$ with an item from it randomly. Set $c = a + b$. If $c \leq 1$, put it in the appropriate range as a new item and remove $a$ and $b$ from their ranges. Go to line 11.

15- If there is at least one item in $(0.1, 0.2)$, initialize pointer $b$ with an item from it randomly. Set $c = a + b$. If $c \leq 1$, put it in the appropriate range as a new item and remove $a$ and $b$ from their ranges. Go to line 11.

16- If there is at least one item in $(0, 0.1)$, initialize pointer $b$ with an item from it randomly. Set $c = a + b$. If $c \leq 1$, put it in the appropriate range as a new item and remove $a$ and $b$ from their ranges. Go to line 11.

17- Put $a$ in a new bin, remove it from its range and go to line 11.

18- If there are not any items in $(0.7, 0.8)$ go to line 24.

19- Initialize pointer $a$ with one of the items from $(0.7, 0.8)$ randomly

20- If there is at least one item in $(0.2, 0.3)$, initialize pointer $b$ with an item from it randomly. Set $c = a + b$. If $c \leq 1$, put it in the appropriate range as a new item and remove $a$ and $b$ from their ranges. Go to line 18.

21- If there is at least one item in $(0.1, 0.2)$, initialize pointer $b$ with an item from it randomly. Set $c = a + b$. If $c \leq 1$, put it in the appropriate range as a new item and remove $a$ and $b$ from their ranges. Go to line 18.

22- If there is at least one item in $(0, 0.1)$, initialize pointer $b$ with an item from it randomly. Set $c = a + b$. If $c \leq 1$, put it in appropriate range as a new item and remove $a$ and $b$ from their ranges. Go to line 18.

23- Put $a$ in a new bin, remove it from its range and go to line 18.

24- If there is not at least one items in $(0.8, 0.9)$ go to line 29.

25- Initialize pointer $a$ with one of the items from $(0.8, 0.9)$ randomly.

26- If there is at least one item in $(0.1, 0.2)$, initialize pointer $b$ with an item from it randomly. Set $c = a + b$. If $c \leq 1$, put it in appropriate range as a new item and remove $a$ and $b$ from their ranges. Go to line 24.

27- If there is at least one item in $(0, 0.1)$, initialize pointer $b$ with an item from it randomly. Set $c = a + b$. If $c \leq 1$, put it in appropriate range as a new item and remove $a$ and $b$ from their ranges. Go to line 24.

28- Put $a$ in a new bin, remove it from its range and go to line 24.

29- If there are not any items in $(0.9, 1)$ go to line 33.

30- Initialize pointer $a$ with one of the items from $(0.9, 1)$ randomly.

31- If there is at least one item in $(0, 0.1)$, initialize pointer $b$ with an item from it randomly. Set $c = a + b$. If $c \leq 1$, put it in the appropriate range as a new item and remove $a$ and $b$ from their ranges. Go to line 29.

32- Put $a$ in a new bin, remove it from its range and go to line 29.

33- If there are not any items in $(0.4, 0.5)$ go to line 35.

34- Initialize pointers $a$ and $b$ with two items from $(0.4, 0.5)$ randomly. Set $c = a + b$. Put it in the appropriate range as a new item and remove $a$ and $b$ from their ranges. Go to line 24.

35- If there is not at least one item in $(0.3, 0.4)$ go to line 37.

36- Initialize pointers $a$ and $b$ with two items from $(0.3, 0.4)$ randomly. Set $c = a + b$. Put it in appropriate range as a new item and remove $a$ and $b$ from their ranges. Go to line11.

37- If there are not any items in $(0.2, 0.3)$ go to line 39.

38- Initialize pointers $a$ and $b$ with items from (0.2, 0.3) randomly. Set $c = a + b$. Put it in appropriate range as a new item and remove $a$ and $b$ from their ranges. If $c \geq 5$, go to line 3 otherwise go to line 33.

39- If there are not any items in $(0.1, 0.2)$ go to line 41.

40- Initialize pointers $a$ and $b$ with items from (0.1, 0.2) randomly. Set $c = a + b$. Put it in the appropriate range as a new item and remove $a$ and $b$ from their ranges. Go to line 33.

41- If there are not any items in $(0, 0.1)$ go to line 43.

42- Initialize pointers $a$ and $b$ with items from $(0, 0.1)$ randomly. Set $c = a + b$. Put it in the appropriate range as a new item and remove $a$ and $b$ from their ranges. Go to line 37.

43- End

The state which only one S item remains in steps 34, 36, 38, 40 and 42 is ignored because of simplification in understanding the algorithm, but definitely, it cannot be ignored in the complete algorithm.

## 3. The Approximation Ratio of the Algorithm

In this section, it is proved that the approximation ratio of the algorithm is 3/2. Firstly, 4 types of errors are introduced which are all possible types of the algorithm errors. These errors are actually the possible differences between the optimal solution and the algorithm's solution. If there are not any errors, the algorithm will produce the optimal solution. Next, the algorithm with only error type1 is discussed and it is proved that the approximation ratio in this condition is 3/2. After that, other types of errors are added one by one until finally, it is proved that if all types of the errors exist, the approximation ratio will be 3/2, again. Note that in all steps we assume that the errors are maximized, and the algorithm is given the worst possible input.

*Algorithm errors:*

1- An L item is matched with an S item which belongs to an S pack (All items in this kind of packs are small).

2- An L item is matched with an S item which belongs to an L pack (a pack at which there is one L item) whereas the L item of the L pack and the first L item belong to the same range.

3- An L item is matched with an S item which belongs to an L pack whereas the L item in this pack belongs to a range except the range correspond to the mentioned L item.

4- Error in matching S items (both in S packs and L packs): S items are matched in a wrong manner; it means that they are matched in a different way from the optimal solution.

The output of the algorithm will be some filled bins which can have two general states which are shown in figure 1. Some of output bins contain just some small items, but others have exactly one large item and maybe some small items.

```
┌──────┬──────┬─────┬──────┐
│ S₁   │ S₂   │ ... │ Sₖ   │
└──────┴──────┴─────┴──────┘
   1 ≤ k ≤ n

┌──────┬──────┬──────┬─────┬──────┐
│ L    │ S₁   │ S₂   │ ... │ S'ₖ  │
└──────┴──────┴──────┴─────┴──────┘
   0 ≤ k' ≤ n − 1
   n: Number of inputs
```

*Figure 1: The states of the output bins in a OPT solution*

Based on the algorithm, an L item begins to pick up an S item. The L item can make a wrong choice and pick up an S item which belongs to an L pack whereas the L item of the L pack and the first L item belong to the same range (error type 2) or the L item of the L pack belongs to ranges except the range of the first L item (error type 3) or pick up an S item which is relevant to an S pack (error type 1). S items also can be matched with each other in a wrong manner (error type 4). The errors are maximized in each step of the proof to engender the worst condition.

**Step 1:** Suppose there is only error type 2. We prove that in this condition $Max\ (P^*/P) = 3/2$.

***Lemma 1:*** In the worst case (the state which there are the most possible differences between the algorithm's output and the OPT solution), there are no bins with only an L item or only an S item in OPT solution.

***Proof:*** We use proof by contradiction; suppose there are some bins with only an L item or only an S item in OPT solution, so the presented algorithm and all other possible algorithms will put this S item (or L item) in a separate bin or will match it with some other items. In both states, the algorithm does not perform worse than OPT solution and therefore in this condition, the problem would be reduced to an easier problem, and obviously it is not the worst case.

If there are $m_1$ bins with only an S item and $m_2$ bins with only an L item, then seemingly the problem is degraded to a problem with $n - m_1 - m_2$ items, therefore we never have bins with only an L or an S item in OPT solution in the worst case. ∎

***Corollary 1***: The output bins in each optimal solution have one of the states which are shown in figure 2.

```
┌──────┬──────┬─────┬──────┐
│ S₁   │ S₂   │ ... │ Sₖ   │
└──────┴──────┴─────┴──────┘
   2 ≤ k ≤ n
(a)

┌──────┬──────┬──────┬─────┬──────┐
│ L    │ S₁   │ S₂   │ ... │ S'ₖ  │
└──────┴──────┴──────┴─────┴──────┘
   1 ≤ k' ≤ n − 1
(b)
   n: Number of inputs
```

*Figure 2: The possible states of a OPT solution which maximize the errors*

In this step, there are only bins which follow the state (b) in figure 2, and the existence of bins like figure 2-(a) is not reasonable because there is not error type 2 in this step. The S packs in the OPT solution and the algorithm's output are the same because in this step, there is no error type 4.

The S item matched with the L item would not be in the lower ranges than the L item's complementary range, otherwise at least one L item would be matched with the S item while we want to maximize error type 2 and send all the L items into bins, alone; therefore, the worst-case in this step will happen when all L items are in the same range and their complementary S items are in complementary range. Therefore, we just have bins which is in the state of figure 2-(b) in this step.

In conclusion, there are $\frac{n}{2}$ L items and $\frac{n}{2}$ S items in a couple of complementary ranges. For example all S items belong to $(0.3, 0.4)$ and all L items belong to $(0.6, 0.7)$. Therefore, all L items are in $(m, m + 0.1)$ range and all S item are in $(1 - m - 0.1, 1 - m)$. On the other hand, there are bins only like figure1. Then, we have $\frac{n}{2}$ L items in (m, m+0.1) and $\frac{n}{2}$ S items in $(1 - m - 0.1, 1 - m)$.

Suppose all L items encounter an S item that the L item cannot be matched with. The worst case in this step happens in this condition because it means all L items are packed alone. Based on the algorithm, S items would be matched two by two in the worst case. Finally we have:

$$P^* = n/2 + n/4 = (3/4) * n$$

$$P = n/2$$

Therefore:

$$P^*/P = \frac{(3/4) * n}{n/2} = 3/2$$

Assume that $P^*$ is the output of the algorithm and $P$ is the optimal answer.

**Step 2:** Suppose there are only error types 2 and error type 3. We prove that in such condition $Max\ (P^*/P) = 3/2$.

***Lemma 2***: If the number of L items is more than the number of S items, then $\frac{P^*}{P} = \frac{3}{2}$.

Suppose there are $n_1$ L items and $n_2$ S items and:

$$n = n_1 + n_2, n_1 > n_2$$

Then $n_1$ is the minimum number of bins in OPT solution because two large items cannot be put in a common bin, then:

$$P \geq n_1$$

In the worst case all L items are packed in distinct bins alone and all S items would be match with each other two by two. In this condition the ratio of $\frac{P^*}{P}$ is less than $\frac{3}{2}$ inasmuch as:

$$\frac{P^*}{P} \leq \frac{P^*}{n_1} \leq \frac{n_1 + \frac{n_2}{2}}{n_1} \leq \frac{n_1 + \frac{n_1}{2}}{n_1} \leq \frac{\left(\frac{3}{2}\right) * n_1}{n_1} \leq \frac{3}{2} \Rightarrow P^*/P \leq 3/2. \blacksquare$$

According to lemma 2, we know that the number of the L items is not more than the S items; therefore the state of $(n/2 - \varepsilon)$ L items and $(n/2 + \varepsilon)$ S items will be studied while $0 \leq \varepsilon \leq n/2$.

According to lemma 1, we do not have an output bin in optimal solution which contains only an L item in the worst case, therefore each L item is matched with one or more S items in OPT solution. The worst case occurs when an L item is matched with only one S item because there is not error type 4 in this step. For instance, if $L_j$ is matched with $s_1, s_2, \ldots, s_i$ items in one bin in OPT solution, then it means $L_j$ is matched with $s' = s_1 + s_2 + \ldots + s_i$ in one bin and the problem is degraded to a problem with $n - i + 1$ inputs. Therefore, $\frac{n}{2} - \varepsilon$ of the L items are matched with S items one by one.

As proved, each L item must be matched with only one S item; then $2\varepsilon$ of the S items will remain. $2\varepsilon$ S items will be matched with each other efficiently because in this step there is no error type 4. In conclusion, the problem is degraded to a problem with $n - 2\varepsilon$ inputs; meaning for maximizing the errors, $\varepsilon$ must be 0.

In the worst case, every L item goes into a bin alone and the S items are matched with each other two by two. On the other hand, it is concluded that $P \geq \frac{n}{2}$ because definitely, each L item requires one bin and there are $\frac{n}{2}$ L items. Therefore:

$$Max\left(\frac{p*}{P}\right) \leq \frac{\frac{n}{2} + \frac{n}{4}}{\frac{n}{2}} = \frac{3}{2}$$

**Step 3:** In this step, the errors type 1, 2, 3 are considered. In this case also the ratio of $P^*/P$ is less than 3/2.

According to lemma 1 and 2, we know final bins in OPT solutions follow figure 2.

Despite various solutions for every problem, here, it is assumed that there is only one OPT solution, therefore the attempts are aimed at getting closer to the only possible final OPT solutions in the presented algorithm. Clearly, it does not confine any general aspect of the problem and will make the problem even harder.

There is no errors in matching $m_2$ S packs because there is no error type 4 in this step. For example, if there are m S items in all S packs, the problem is reduced to a problem with $n - m$ items. In conclusion, $m_2$ is considered as 0 in this step.

According to lemma 1, there is not an L item alone in output in the worst case, therefore each L item is matched with one or more S items in OPT solution. The worst case occurs when an L item is matched with only one S item in that there is no error type 4 in this step. For example, if $L_j$ is matched with $s_1, s_2, \ldots, s_i$ items in one bin in OPT solution, it means Lj is matched with $s' = s_1 + s_2 + \ldots + s_i$ in one bin and the problem is transformed into an easier problem with $n - i + 1$ inputs.

It is supposed that all L items go into distinct bins alone. It is obvious that the worst case occurs in this condition because if any S item would be matched with the L item, the problem would be reduced to an easier problem and it is not the worst case.

Now, when $m_1$ L item goes into bins alone then S items will be matched with each other two by two. Therefore we have:

$$Max\left(\frac{P^*}{P}\right) = \frac{m1 + \frac{m1}{2}}{m1} = \frac{\left(\frac{3}{2}\right)m1}{m1} = \frac{3}{2}$$

**Step 4:** Now, the state in which there is only error type 4 is discussed. The L items can be ignored because the only existing error would be the type 4 and the L items cannot engender any errors in this step. It means that there are only S items.

In this step, it is claimed that the worst state will happen when all items are in $(0.3, 0.4)$. Suppose that items have been distributed arbitrarily in all ranges. Based on the algorithm, first, the items that are in $(0.4, 0.5)$ will be coupled with each other two by two and will create a new item existing between 0.8 and 1.0. After that, this new item will try to be matched with the items located in $(0, 0.1)$ or $(0.1, 0.2)$. After that, each of the final created item will be put in one bin. Finally, at least 0.8 size of each output bin is full.

Based on the algorithm, after finishing the items in $(0.4, 0.5)$, items in $(0.3, 0.4)$ will be joined together in binary and triad groups, respectively and will create the new item in between 0.6 and 1.0. After that, these new item will be matched with the items existing between 0 and 0.4. Each of the final created item will be put in one bin. Finally, at least 0.6 of each output bin's size is full.

Definitely, after finishing the items in $(0.3, 0.4)$, the items relevant to $(0.2, 0.3)$ will be matched with each other. In this condition, there are not any items in $(0.3, 1.0)$. Two item from $(0.2, 0.3)$ will be matched with each other and will create a new item between 0.4 and 0.6. After that, the new created item will be matched with an item in $(0.2, 0.3)$ and will make a new item. This process continues which in consequence the created item becomes larger and larger. Finally, at least 0.7 of each output bin's size is full because the biggest item of Small set is 0.3.

Absolutely, after finishing the items in $(0.2, 0.3)$, there are not any items between 0.2 and 1.0. Obviously, with mentioned proof for $(0.2, 0.3)$ at least 0.8 size of each output bin is full because the biggest item of Small set is 0.2 in this step.

In next step, the only remaining items will be in $(0, 0.1)$. Definitely, at least 0.9 of each output bin's size is full because the biggest item of Small set is 0.1.

***Lemma 3***: Definitely, $\frac{P^*}{P} \leq \frac{3}{2}$ if at least $\frac{2}{3}$ size of each output bin is full.

***Proof***: Suppose the worst condition (all of output bins are completely full in OPT solution) and $W$ is the sum of input items. In this condition, we have:

$$P \geq W \ \& \ P^* \leq \frac{W}{\frac{2}{3}} \Rightarrow \frac{P^*}{P} \leq 3/2 \ \blacksquare$$

According to lemma 3, items in ranges $(0, 0.1)$, $(0.1, 0.2)$, $(0.2, 0.3)$ and $(0.4, 0.5)$ are not the problem because at least 0.7 of their final output bins' size are full. On the other hand, $(0.3, 0.4)$ could cause some problems because it can create bins that are smaller than $2/3$.

Therefore, for creating the worst condition in this step, we will put all the items in $(0.3, 0.4)$ because the total free space in output bins and the number of output bins in the algorithm will be maximized.

The proportion of P* to P is not going to be more than $3/2$ in this step because final bins in OPT solution cannot have more than 3 items, and on the other hand, S items at least will be matched with each other two by two based on the algorithm. Therefore the algorithm will have maximum proportion of $3/2$ in this condition. Therefore:

$$n = 3m \Rightarrow P^* = \frac{3m}{2}, P = m \Rightarrow Max\left(\frac{P^*}{P}\right) = \frac{3}{2}$$

**Step 5:** The final and general state which contain all kinds of errors is discussed, it is proved that that $MAX\ (P^*/P)$ is not more than 3/2, and therefore, the approximation ratio is 3/2.

***Theorem 1***: The proposed algorithm is a $\frac{3}{2}$-approximation algorithm.

Based on Lemma 1 and Lemma 2, it is acknowledged that output bins in OPT state are like figure 2.

We suppose that $m_1 = m'_1 + m''_1$.

$m'_1$= the number of L items alone in a bin in OPT state.
$m''_1$=the number of L items that are at least with an S item in OPT state.

Obviously, in the general state the size of $m'_1$ and $m''_1$ are arbitrary. It is proved that in this condition $Max\ (P^*/P)$ will not be more than 3/2, again.

First, the state in which each L item is in one bin alone in OPT solution will be discussed. Based on step 2, each L item will be put in one bin alone, if the $(n-z)/2$ L item are in the same range $(m, m + 0.1)$ and $(n-z)/2$ S item are in its complementary range $(1 - m - 0.1, 1 - m)$. On the other hand, S items that are in S packs must belong to higher ranges otherwise, at least there is an L item that will be put in a bin with an S item.(z is equal to the number of S item in S packs)

Obviously, we want to maximize the number of output bins for the presented algorithm. It is obvious that the L items need $m_1$ bins and it is demanded to maximize the number of bins that S items (S item that are in S packs and L packs) need, based on the algorithm. Definitely, these S items will produce the most number of output bins if there is the maximum free space in bins and based on step 4, it is acknowledged that the best range for the mentioned condition is $(0.3, 0.4)$. In conclusion, in the worst condition there will be $m_1$ S items that belong to L packs and $3m_2$ S items that belong to S packs. The maximum number of S items in $(0.3, 0.4)$ is $3m_2 + m_1$. Definitely, these S items will be matched with each other 2 by 2. Then:

$$Max\ (P^*) = \frac{3m_2 + m_1}{2} + m_1\ \&\ Min\ (P) = m_1 + m_2 \Rightarrow Max\ (P^*/P) = \frac{\frac{3m_1}{2} + \frac{3m_2}{2}}{m_1 + m_2} = 3/2$$

Now, consider general state in which there are $m_1 = m'_1 + m''_1$, L packs and $m_2$ S packs. Based on 2, $m'_1$ L item must belong to the same range like $(m, m + 0.1)$ and the S items in these packs must belong to complementary range $(0.9 - m, 1.0 - m)$. Other S items must belong to higher ranges otherwise they will be matched with some $m'_1$ L items and it is in contrast with the assumption. Furthermore, other L items must belong to the lower ranges than the mentioned range (the range that is related to $m'_1$ L item) otherwise they will not be matched with any S items and it is in contrast with the assumption, again.

$m'_1$ L items would belong to five different ranges $(0.5, 0.6), (0.6, 0.7), (0.7, 0.8), (0.8, 0.9), (0.9, 1.0)$ therefore there are five states. It is going to be proved that in all these states $Max\ (P^*/P)$ will not be more than 3/2.

*State 1*:

The distribution of the input items follow figure 3 in state 1.

| 5 – 6 | 6 - 7 | 7 – 8 | 8 – 9 | 9 – 10 |
|---|---|---|---|---|
| ■ |  |  |  |  |

| 4 – 5 | 3 – 4 | 2 – 3 | 1 – 2 | 0 – 1 |
|---|---|---|---|---|
| ■ |  |  |  |  |

*Figure 3: The ranges that can contain some items in state 1 are shown by black*

$m'_1$ L items belong to $(0.5, 0.6)$ and their complementary S items belong to $(0.4, 0.5)$. It was discussed before in this study that other S items cannot belong to lower ranges and other L items cannot belong to higher ranges therefore all S items must be put in $(0.4, 0.5)$ and all L item must be put in $(0.5, 0.6)$; therefore:

$$P = m_1 + m_2 = m'_1 + m''_1 + m_2$$

$$P^* = m'_1 + m''_1 + \frac{2m_2 + m'_1 + m''_1 - m''_1}{2} = m'_1 + m''_1 + m_2 + \frac{1}{2}m'_1 \Rightarrow Max\ (P^*/P) = 3/2$$

*State 2*:

| 5 – 6 | 6 – 7 | 7 – 8 | 8 – 9 | 9 – 10 |
|---|---|---|---|---|
|  | ■ |  |  |  |

| 4 – 5 | 3 – 4 | 2 – 3 | 1 – 2 | 0 – 1 |
|---|---|---|---|---|
|  | ■ |  |  |  |

*Figure 4: The ranges that determine the boundary which the inputs items can be put in state 2*

$m'_1$ L items belong to $(0.6, 0.7)$ and their complementary S items belong to $(0.3, 0.4)$. It has been discussed in this article that other S items cannot belong to lower ranges and other L items cannot belong to higher ranges therefore other S item will be put in $(0.3, 0.5)$ and other L items will be put in $(0.5, 0.7)$. To simplify, in this section's figures we use $i$ instead of $0.i$ to show the boundaries of the ranges.

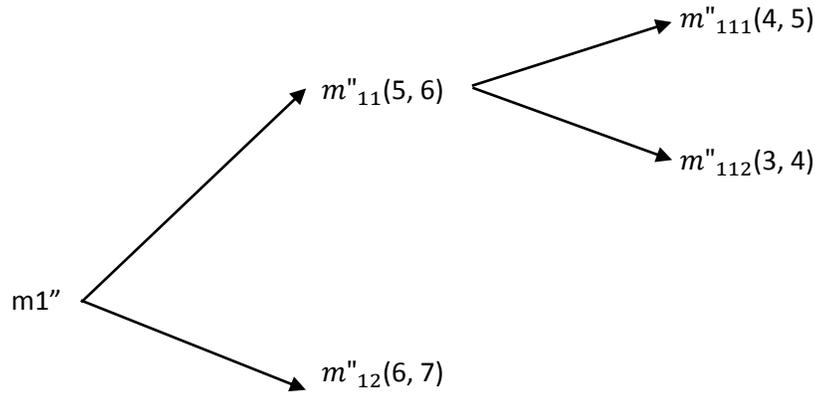

The range that is in front of $m"_{ij}$ shows its own range and the range in front of $m"_{ijk}$ shows the $k^{th}$ possible complementary range or ranges for $m"_{ij}$.

For example, in this case $m"_1$ L items are related to $(0.5, 0.6)$ or $(0.6, 0.7)$ that are named $m"_{11}$ and $m"_{12}$, respectively. The L items in $(0.5, 0.6)$ as it had been previously mentioned are showed with $m"_{11}$. There are some S items that are in the same L pack with these items in OPT solution. These S and L items can be put in output bins in different ways like one L item in $(0.5, 0.6)$ and one S item in $(0.4, 0.5)$ or one L item in $(0.5, 0.6)$ and one S item in $(0.3, 0.4)$ that are named $m"_{111}$ and $m"_{112}$, respectively.

The $m"_{112}$ can be ignored because $m"_{111}$ includes it. In each combination that an item in $(0.4, 0.5)$ can be put, an item in $(0.3, 0.4)$ also can be put while an item in $(0.4, 0.5)$ can make more difficulties in matching because of its size since it is desirable to engender the worst case. Now we have:

$$P = m_1 + m_2 = m'_1 + m"_1 + m_2 = m'_1 + m"_{11} + m"_{12} + m_2$$
$$P^* = m'_1 + m"_1 + \frac{3m_2 + m'_1 + m"_1 - m"_1}{2} \Rightarrow Max\left(\frac{P^*}{P}\right) = \frac{3}{2}$$

*State 3*:

| 5 – 6 | 6 – 7 | 7 – 8 | 8 – 9 | 9 – 10 |
|---|---|---|---|---|
|  |  | ▓ |  |  |

| 4 – 5 | 3 – 4 | 2 – 3 | 1 – 2 | 0 – 1 |
|---|---|---|---|---|
|  |  | ▓ |  |  |

*Figure 5: The ranges that determine the boundary which the inputs items can be put in state 3*

$m'_1$ L items belong to $(0.7, 0.8)$ and their complementary S items belong to $(0.2, 0.3)$. It is acknowledged that other S items would belong to $(0.2, 0.5)$ and other L items would belong to $(0.5, 0.8)$.

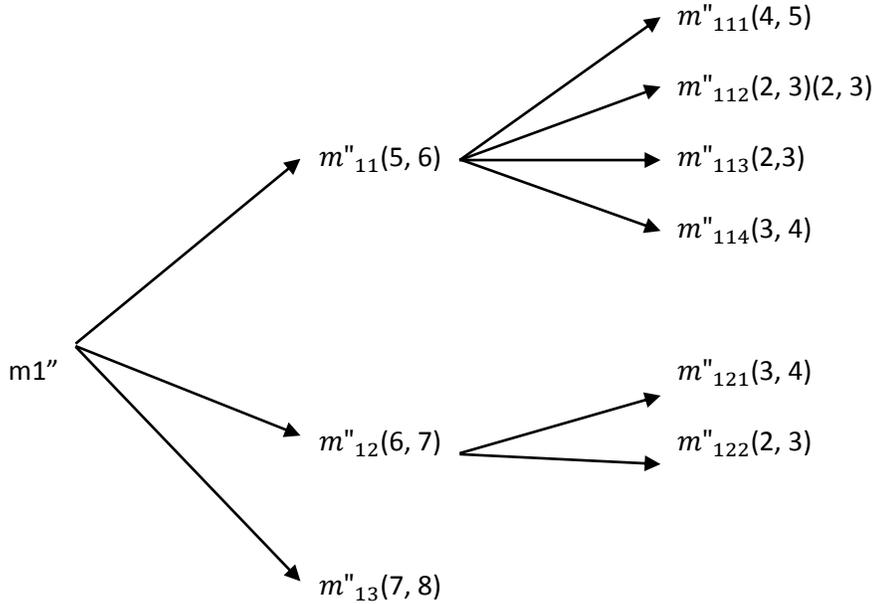

The $m"_1$ L items in $(0.5, 0.8)$ are separated into 3 groups:

$m"_1$: The L items in $(0.5, 0.6)$
$m"_{12}$: The L items in $(0.6, 0.7)$
$m"_{13}$: The L items in $(0.7, 0.8)$

And the $m"_{11}$ L items also are separated into 4 groups:

$m"_{111}$: The L item matched with one S item in $(0.4, 0.5)$ in OPT solution
$m"_{112}$: The L item matched with two S item in $(0.2, 0.3)$ in opt solution
$m"_{113}$: The L item matched with one S item in $(0.2, 0.3)$ in OPT solution
$m"_{114}$: The L item matched with one S item in $(0.3, 0.4)$ in OPT solution

$m"_{113}$ and $m"_{114}$ can be ignored because $m"_{111}$ includes them as well. In each combination that an item belonging to $(0.4, 0.5)$ can be put in, an item in $(0.2, 0.3)$ and an item in $(0.3, 0.4)$ also can be put in while an item in $(0.4, 0.5)$ can engender more obstacles in matching because of its size.

These procedures and reasons for other parameters in this step and step 4 and 5 are similar and they will be ignored for the matter of discussion.

Actually $m"_{112}$ means $m"_1$ which belongs to $(0.5, 0.6)$ and in optimum solution will be matched with two elements of $(0.2, 0.3)$. $m"_{113}$, $m"_{114}$, and $m"_{112}$ are meaningless because as it was mentioned before, wherever item in $(0.4, 0.5)$ could be put in, the item in $(0.3, 0.4)$ could be put in, too. Therefore it would be better for items in $(0.5, 0.6)$ to be matched with an items in $(0.4, 0.5)$. Henceforth meaningless states are ignored.

***Lemma 4:*** The worst case occurs when every large element of L items belongs to $m"_1$ would choose an element from a small range which is not its complementary range.

Based on the algorithm, each item in $m"_1$ has to be matched with an S item. If the chosen S item belongs to a range lower than its complementary one, the worst case becomes more approachable because wherever an item from complementary range could be placed in, an item from mentioned S could be placed in, too.

According to lemma 4, complementary items for elements of $m"_1$ from lower ranges are chosen.

$$p = m_1 + m_2 = m'_1 + m"_1 + m_2 = m"_{111} + m"_{112} + m"_{12} + m"_{13}$$

$$P^* = m'_1 + m"_1 + x$$

| 4 – 5 | 4 – 3 | 3 – 2 |
|---|---|---|
| $m"_{111}$ | $m"_{121}$ | $2*m"_{112}$ $m'_1$ $m"_{13}$ |

In $(0.4, 0.5)$ there are $m"_{111}$ S items. In $(0.3, 0.4)$ there are $m"_{112}$ S items. In $(0.2, 0.3)$ there are $2 * m"_{112} + m'_1$ S items except the S items in S packs.

$$x = \frac{m"_{111}}{2} + \frac{m"_{121} - m"_{11}}{2} + \frac{2*m"_{112} + m'_1 - m"_{12} - m"_{13} + m"_{13}}{3}$$

$$= \frac{3*m"_{111} + 4*m"_{112}}{6} + \frac{m"_{12}}{2} - \frac{m"_{12}}{3} - \frac{m"_{11}}{2} + \frac{m'_1}{3} + \frac{m"_{112}}{6} + \frac{m"_{12}}{6} + \frac{m'_1}{3} + \frac{3*m_2}{2}$$

Obviously, in the worst case, $m_2$ items have to produce maximum bins, so according to the state 4, they are put in $(0.3, 0.4)$.

$$P* = m'_1 + m"_1 + \frac{m"_1}{6} + \frac{m'_1}{3} + \frac{3}{2}m_2 = \frac{4}{3}m'_1 + \frac{7}{6}m"_1 + \frac{3}{2}m_2$$

$$Max\ (P^*/P)\ \leq 3/2$$

*State 4:*

| 5 – 6 | 6 – 7 | 7 – 8 | 8 – 9 | 9 – 10 |
|---|---|---|---|---|
|  |  |  | ▓ |  |

| 4 – 5 | 3 – 4 | 2 – 3 | 1 – 2 | 0 – 1 |
|---|---|---|---|---|
|  |  |  | ▓ |  |

*Figure 6: The ranges that determine the boundary which the inputs items can be put in state 4*

$m'_1$ L items and their complementary S items are in $(0.8, 0.9)$ and $(0.1, 0.2)$.

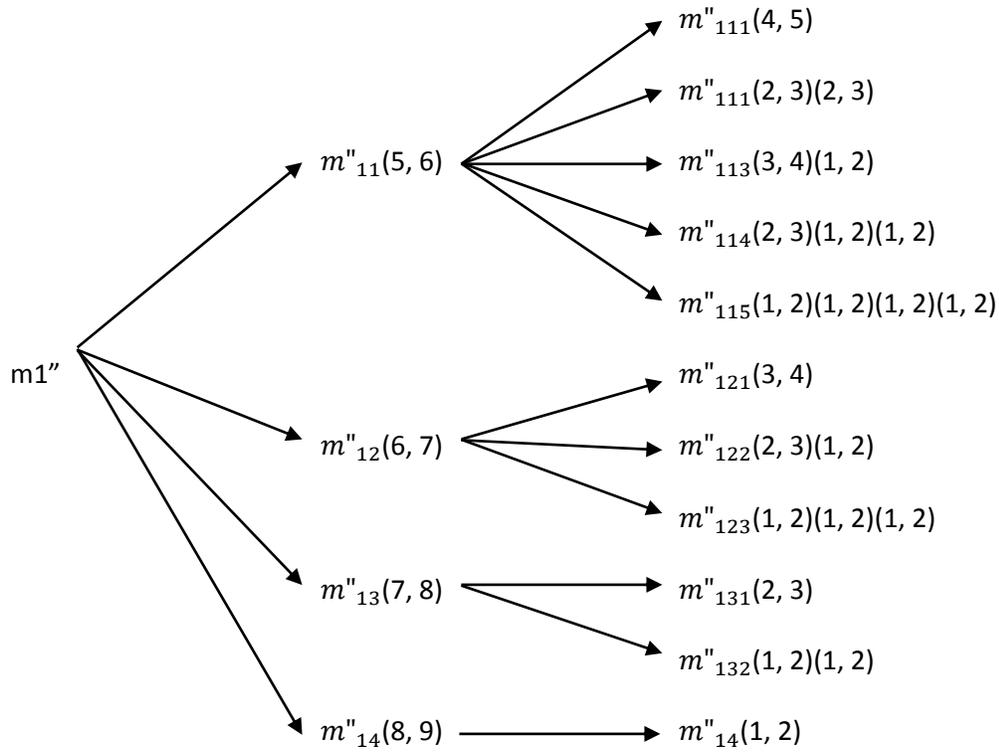

$P = m'_1 + m''_1 + m_1$

$P^* = m'_1 + m'_2 + x$

| 4 – 5 | 3 – 4 | 2 – 3 | 1 - 2 |
|---|---|---|---|
| $m'_{111}$ | $m'_{113}$<br>$m''_{121}$ | $2 * m''_{112}$<br>$m''_{114}$<br>$m''_{122}$<br>$m''_{131}$ | $m''_{113}$<br>$2 * m''_{114}$<br>$4 * m''_{115}$<br>$m''_{122}$<br>$3 * m''_{123}$<br>$2 * m''_{132}$<br>$m''_{14}$<br>$m'_1$ |

$$z = \frac{m"_{111}}{2} + \frac{m"_{113} + m"_{121} - m"_{111}}{2} + \frac{2*m"_{112} + m"_{114} + m"_{122} + m"_{131} - m"_{12}}{3}$$
$$+ \frac{2*m"_{132} + m"_{14} + m'_1 - m"_{13} - m"_{14} + m"_{113} + 2m"_{114} + 4m"_{115}}{5}$$
$$+ \frac{m"_{122} + 3m"_{123}}{5}$$

$$= \frac{15m"_{111} + 20m"_{112} + 21m"_{113} + 22m"_{114} + 24m"_{115}}{30} - \frac{m"_{11}}{2}$$
$$+ \frac{15m"_{121} + 16m"_{122} + 18m"_{123}}{30} - \frac{m"_{12}}{3} + \frac{10m"_{131} + 12m"_{132}}{30} - \frac{3}{5} + \frac{m'_1}{5}$$

$$\Rightarrow z = \frac{5m"_{112} + 6m"_{113} + 7m"_{114} + 9m"_{115}}{30} + \frac{5m"_{121} + 6m"_{122} + 8m"_{123}}{30} + \frac{4m"_{131} + 6m"_{132}}{30}$$
$$+ \frac{m'_1}{5}$$

According to state 4, in the worst case, $m_2$ packs would be placed in $\frac{3m_2}{2}$ bins.

$$x = z + \frac{3}{2}m2$$

$$\max(x) = \max(z) + \frac{3}{2}m_2 = \frac{9}{30}m"_1 + \frac{m'_1}{5} + \frac{3}{2}m_2$$

$$\frac{P^*}{P} = \frac{\frac{3}{10}m_1 + \frac{1}{5}m_1 + \frac{3}{2}m_2 + m'_2 + m"_1}{m1' + m1" + m2}$$

$$\Rightarrow \max\left(\frac{P*}{p}\right) = \frac{3}{2}$$

*State 5*:

| 5 – 6 | 6 - 7 | 7 – 8 | 8 – 9 | 9 - 10 |
|---|---|---|---|---|
|  |  |  |  | ▓ |

| 4 – 5 | 3 – 4 | 2 – 3 | 1 – 2 | 0 – 1 |
|---|---|---|---|---|
|  |  |  |  | ▓ |

*Figure 7: The ranges that determine the boundary which the inputs items can be put in state 5*

L items and their complementary S items are in (0.9, 1) and (0, 0.1).

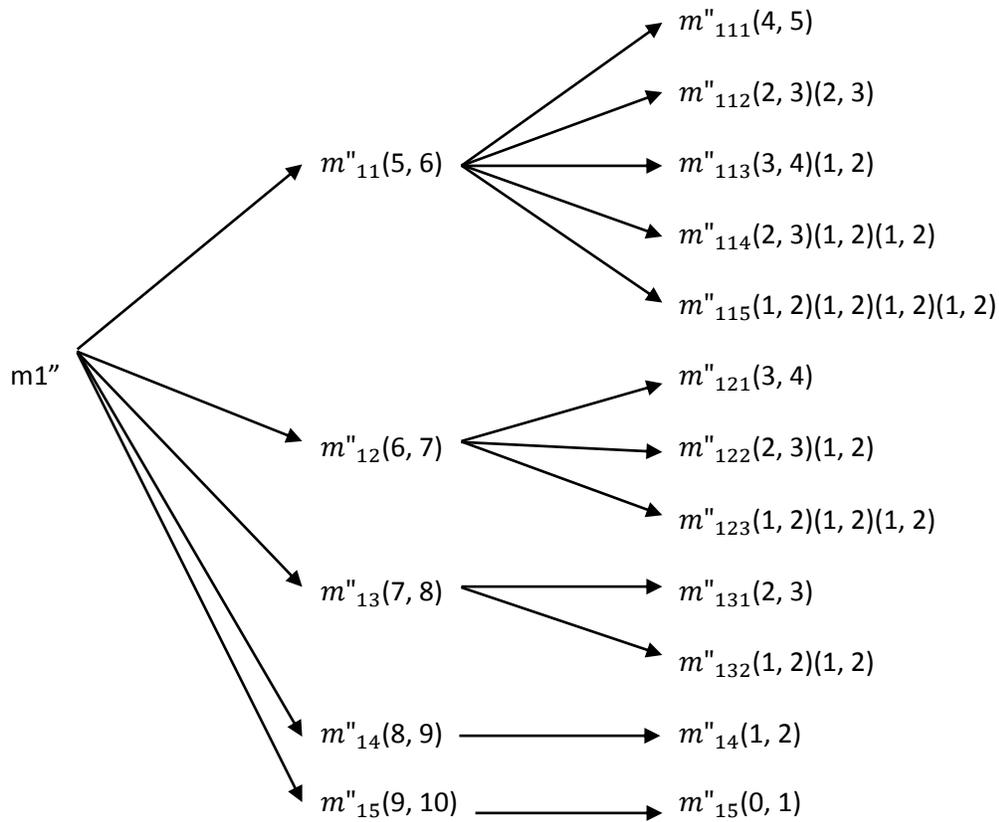

$$P = m'_1 + m''_1 + m_2$$
$$P^* = m'_1 + m''_1 + x$$
$$x = z + \frac{3}{2} m_2$$

| 4 – 5 | 3 – 4 | 2 – 3 | 1 - 2 | 0 - 1 |
|---|---|---|---|---|
| $m''_{111}$ | $m''_{113}$ $m''_{121}$ | $2 * m''_{112}$ $m''_{115}$ $m''_{122}$ $m''_{131}$ | $m''_{113}$ $4 * m''_{114}$ $2 * m''_{115}$ $m''_{122}$ $3 * m''_{123}$ $2 * m''_{132}$ $m''_{14}$ | $m''_{15}$ $m'_1$ |

$$z = \frac{m''_{111}}{2} + \frac{m''_{113} + m''_{121} - m''_{11}}{2} + \frac{2m''_{112} + m''_{115} + m''_{122} + m''_{131} - m''_{12}}{3}$$
$$+ \frac{m''_{113} + 4m''_{114} + 2m''_{115} + m''_{122} + 3m''_{123} + 2m''_{132} + m''_{14} - m''_{13}}{5}$$
$$+ \frac{m''_{15} + m'_1 - m''_{14} - m''_5}{10}$$

$$= \frac{m''_{111}}{2} + \frac{2}{3}m''_{112} + \frac{7}{10}m''_{113} + \frac{4}{5}m''_{114} + \frac{11}{15}m''_{115} - \frac{1}{2}m''_{11} + \frac{1}{2}m''_{121} + \frac{8}{15}m''_{122} + \frac{3}{5}m''_{123}$$
$$- \frac{1}{3}m''_{12} + \frac{1}{3}m''_{131} + \frac{2}{5}m''_{132} - \frac{1}{5}m''_{13} + \frac{1}{10}m''_{14} + \frac{1}{10}m'_1$$

$$= \frac{15m''_{111} + 20m''_{112} + 21m''_{113} + 24m''_{114} + 22m''_{115}}{30} - \frac{m''_{11}}{2} + \frac{15m''_{121} + 16m''_{122} + 18m''_{123}}{30}$$
$$- \frac{m''_{12}}{3} + \frac{5m''_{131} + 6m''_{132}}{15} - \frac{m''_{13}}{5} + \frac{m''_{14}}{10} + \frac{m'_1}{10}$$

$$= \frac{5m''_{112} + 6m''_{113} + 9m''_{114} + 7m''_{115}}{30} + \frac{5m''_{121} + 6m''_{122} + 8m''_{123}}{30} + \frac{4m''_{131} + 6m''_{132}}{30} + \frac{m''_{14}}{10}$$
$$+ \frac{m'_1}{10}$$

$$\max(z) = \frac{3}{10}m''_1 + \frac{1}{10}m'_1$$

$$\max(P^*) = \frac{3}{10}m''_1 + \frac{m'_1}{10} + m''_1 + m'_1 + \frac{3}{2}m_2$$

$$\max\left(\frac{P^*}{P}\right) = \max\left(\frac{\frac{13}{10}m''_1 + \frac{11}{10}m'_1 + \frac{3}{2}m_2}{m'_1 + m''_1 + m2}\right) = \frac{3}{2}$$

In conclusion, $\frac{P*}{P}$ never exceeds 3/2; therefore the approximation ratio of the suggested algorithm is 3/2.
∎

## 6. Conclusion

Bin Packing Problem is an important problem which is used in various fields. Since this problem is an NP-hard problem, researchers have been trying to solve it with the approximation approaches. It is proved that the best approximation factor and the best time order for this problem are 3/2 and $O(n)$, respectively. In this article, a new algorithm for this problem was presented which can produce a result in $O(n)$, and it was proved that the approximation factor for this problem is 3/2. In this proof, four kinds of errors were considered which are the only possible errors in the algorithm, and it is tried to maximize the possible errors step by step. Finally, it was proved the approximation factor in all conditions for the presented algorithm is 3/2.